# Hierarchical self-organization of cytoskeletal active networks

Daniel Gordon[1], Anne Bernheim-Groswasser[2,5],
Chen Keasar[1,3], Oded Farago[4,5]

Departments of [1]Computer Science, [2]Chemical Engineering, [3]Life Sciences, [4]Biomedical Engineering, and [5]Ilse Katz Institute for Nanoscale Science and Technology, Ben Gurion University of the Negev, Be'er Sheva 84105, Israel.

## Abstract

The structural reorganization of the actin cytoskeleton is facilitated through the action of motor proteins that crosslink the actin filaments and transport them relative to each other. Here, we present a combined experimental-computational study that probes the dynamic evolution of mixtures of actin filaments and clusters of myosin motors. While on small spatial and temporal scales the system behaves in a very noisy manner, on larger scales it evolves into several well distinct patterns such as bundles, asters, and networks. These patterns are characterized by junctions with high connectivity, whose formation is possible due to the organization of the motors in "oligoclusters" (intermediate-size aggregates). The simulations reveal that the self-organization process proceeds through a series of hierarchical steps, starting from local microscopic moves and ranging up to the macroscopic large scales where the steady-state structures are formed. Our results shed light into the mechanisms involved in processes like cytokinesis and cellular contractility, where myosin motors organized in clusters operate cooperatively to induce structural organization of cytoskeletal networks.

PACS

87.16.Ka    Filaments, microtubules, their networks, and supramolecular assemblies

87.16.Nn    Motor proteins (myosin, kinesin dynein)

87.16.Uv    Active transport processes

87.16.A-    Theory, modeling, and simulations

# 1. Introduction

Living cells need to impose inner order, transport organelles from site to site, withstand external pressures and propagate themselves to different locations [1]. In all of these processes, the cell cytoskeleton is a key player. The cytoskeleton is an out-of-equilibrium 3D network of polar elastic filaments that constantly remodels. This is achieved via a large number of associated proteins that regulate the rates of assembly and disassembly of the cytoskeletal filaments (treadmilling) [2]. The dynamics of the cytoskeleton is also governed by the action of motor proteins. These molecular machines convert chemical energy into mechanical work to generate driving forces and movement [3]. Motor proteins that move along polar filaments are used to transport cargos in cells [4]. Motors may also serve as (active) linkers between cytoskeletal filaments, which give rise to complex structural and dynamical self-organization phenomena [5-7].

Because of the plethora of factors and cellular components involved in cytoskeletal organization, much of our current knowledge of the associated processes comes from controlled, in vitro, experiments of simplified reconstituted model systems [8]. Such simple model systems, consisting of elastic filaments and their associated molecular motors, can exhibit a rich variety of structural patterns including asters, vortices, rings, bundles, and networks [5,6,9,10]. From a theoretical perspective, it is impossible to model these structurally complex systems in full atomistic detail. The existing models, therefore, are based on phenomenological descriptions which address the problem on larger length- and time-scales. Earlier models were based on the introduction of continuum mean field kinetic equations to describe the dynamics of filaments moving relatively to each other due to the presence of cross-linking motors [11-13]. More recently, a new approach was proposed, treating filaments-motors systems as a viscoelastic polar active gel [14,15]. In these generalized hydrodynamic theories, the dynamics is inferred from symmetry considerations or by coarse-graining the mesoscopic kinetic equations. Several inhomogeneous structures have been identified as steady-state solutions of the macroscopic equations, including asters, vortices and spirals [16-19]. The same structures have also been predicted through a somewhat different approach involving coupled dynamical equations for the filaments orientation field and motors density [20,21].

Coarse-grained (CG) simulations that use simplified representations of the participating molecules and the interactions between them have also been employed for studying the dynamics of active gels. Nedelec and co-workers used such simulations to investigate the dynamics in systems consisting of microtubules (MTs) and kinesin-like motors [22,23]. In those simulations the MTs were represented as inextensible elastic polar rods and motors were modeled as small mechanical machines that walk over the MTs, bind them to each other, and lead to their relative movement. Using this model they managed to reproduce asters and vortices, study aster formation dynamics and examine the effects of changing the probability of motor disconnection from the end of a filament. More recently, Head *et al.* studied the active self-organization of motors-filaments systems using a CG model that also takes into account the excluded volume interactions between the polar filaments [24]. Due to the excluded volume effects, the polar filaments exhibit a nematic ordering that breaks down (albeit not into aster-like structures) at high motor densities. Aster formation has been observed in simulations of three-dimensional elastic networks with cross-links (implicitly representing the motors) that can break and reform [25].

In this paper we present a combined computational and experimental study of the self-organization behavior of actin filaments driven by myosin II molecular motors. While in all other previously studied model systems the motor units acted as cross-linkers between two filaments, here we consider a very different scenario. The myosin II motors in our model system are organized in mesoscopic clusters of several (typically 10-20) motors. By grouping several highly non-processive myosin II motors into clusters, one generates processive elements useful for cytoskeleton remodeling. We use fluorescence microscopy and Molecular Dynamics simulations to investigate the self-organization of polar actin filaments by such motor clusters. Our simulation results, which agree well with the experimental observations, provide a molecular level picture of how the motors are involved in the association, binding, and transport of filaments. Our results reveal very noisy dynamics on the small spatial- and temporal-scales, which is attributed to the non-processive nature of individual myosin II motors. Nevertheless, the outcome of these stochastic dynamics is a set of well-characterized steady-state structures like bundles, asters, and cross-linked networks, at larger scales.

## 2. Materials and Methods

### 2.1 Experimental setup

Actin was purified from rabbit skeletal muscle acetone powder [26]. Purification of myosin II skeletal muscle is done according to standard protocols [27]. Recombinant fascin [28] was expressed in *E. coli* as a GST fusion protein. Actin was labeled on Cys374 with Alexa 568 (Invitrogen).

The motility medium contained 10 mM HEPES, pH = 7.67, 1.7 mM Mg-ATP, 5.5 mM DTT, 0.12 mM Dabco (1,4 diazabicyclo[2,2,2]octane), 0.13M KCl, 1.6mM $MgCl_2$, 1% BSA, an ATP regenerating system 0.1mg/ml Creatine Kinase and 1mM Creatine Phosphate, 20μM of G-actin, and various concentrations of myosin II and fascin.

Actin assembly was monitored by fluorescence using an Olympus 71X microscope (Olympus Co., Japan). The labeled actin fraction was 0.04 and the temperature of the experiments was 22°C. Time-lapse images were acquired using an Andor DV887 EMCCD camera (Andor Co., England). Data acquisition and analysis was performed using METAMORPH (Universal Imaging Co.). To prevent protein adsorption, the glass coverslips were coated with an inert polymer (PEG-mal, Nectar Co.) according to a standard protocol [29].

### 2.2 Computer model and simulations

In our coarse-grained model, only the actin and myosin are simulated explicitly. The other components will be considered only implicitly by their effect on the system's behavior. The presence of solvent is implicated by allowing non-bound myosin motors to diffuse and by over-damping force-driven movements. ATP hydrolysis is represented by the ability of myosin motors to move along filaments which they are connected to. The presence of fascin is represented by rendering the actin filaments as stiff as bundles of fascin linked filaments. Thus, we represent fascin linked bundles of actin filaments as single units. Along this manuscript we regard them as *"filaments"* to avoid confusion with the dynamic, ATP-hydrolysis driven, bundles whose formation is an emergent characteristic of our system (see below). Single actin filaments (which are the constituents of our "filaments") will be referred to as F-actin. Similarly, aggregates of myosin motors are represented as single motor units, with the number of motor heads and the total length of the motor aggregate defined by parameters. Finally, we have decided to simulate the actin-myosin system as pseudo two dimensional, which implies that our model does not include any excluded volume effect. This way, the common situation of two filaments lying one on top of the other is represented by two crossing filaments that do not interact.

Actin filaments are represented by linear chains of $N+1$ nodes, with one of the end nodes defined as the "plus" end and the other as the "minus" end. Each pair of neighboring nodes represents a segment of the filament with predefined length $l_{f,0}$. The geometry and elastic properties of the filaments are governed by two energy terms: The first term applies a Hookean spring between each pair of neighboring nodes, keeping their mutual distance close to the predefined rest distance.

$$E_1 = \frac{1}{2} k_f \left( |x_{i+1} - x_i| - l_{f,0} \right)^2, \qquad (1)$$

where $k_f$ is the filaments spring constant and $x_i$ is the coordinate of the i-th node (vector notation is omitted for brevity). The second energy term, which represents the bending rigidity of the filament, assign the following energy term with each node (except for the two edges nodes)

$$E_2 = \frac{1}{2} A (2x_i - x_{i+1} - x_i)^2, \qquad (2)$$

where $A$ is related to the filament's persistence length, $\xi$, by $A = (\xi k_B T)/2(l_{f,0})$. The total energy of the filament is the sum of the above two terms for all the segments and nodes, and the associated force acting on the i-th node is calculated by $f_i = -\partial E / \partial x_i$.

In our model, actin filaments do not interact directly, but rather affect each other through the forces generated by the motors that connect them. The motors are grouped in clusters, each of which includes $2n_h = 20$ individual motors (referred to as "motor heads"). The motor cluster is represented by a long rod whose length $l_m$ is governed by an elastic energy of similar form (but with different parameters) to Eq. (1)

$$E_1^m = \frac{1}{2} k_m (l_m - l_{m,0})^2. \qquad (3)$$

$n_h$ motor heads emanate from each end of this rod. The motor heads stochastically bind to nearby filaments, with each motor head connected to no more than one filament at a time and no more than $\max_h$ heads from the same motor cluster connected to the same filament. Motor heads, even from the same end of the cluster, can bind simultaneously to different actin filaments. Once the motor is connected, it acts like a Gaussian spring (similar to a Hookean spring, but with no rest length) with spring constant $k_h$, and generates a force between the (end of the) motor cluster and the attachment point on the filament. Since the filament is represented as a chain of nodes, the force $\vec{F}$ on the filament does not act at the attachment point $x$, but rather is split between nodes $x_i$ and $x_{i+1}$ on both sides of the attachment point ($x_i \leq x \leq x_{i+1}$). The force on each node is given by the lever rule [23]: $\vec{F}_i = \vec{F} \left| \frac{x - x_{i+1}}{x_i - x_{i+1}} \right|$ and $\vec{F}_{i+1} = \vec{F} \left| \frac{x - x_i}{x_i - x_{i+1}} \right| = \vec{F} - \vec{F}_i$ .

Once connected, a motor head starts advancing towards the filament's plus end. A motor head has a characteristic speed $v_0$, which is the speed at which a motor progresses along a filament in the absence of an externally applied force. While moving along the filament, the length of the Gaussian spring connecting the motor head to the associated motor cluster changes, which leads to changes in the force applied on the motor head. We employ the commonly used linear relationship between the moving velocity and the projection of the applied force along the direction of motion, $F_{axis} = \vec{F} \cdot (\vec{x}_{i+1} - \vec{x}_i) / |\vec{x}_{i+1} - \vec{x}_i|$ (see, e.g., [23]):

$$v = \begin{cases} 0 & F_{axis} \leq -f_0 \\ v_0 \left(1 + F_{axis}/f_0\right) & -f_0 < F_{axis} < f_0 \\ 2v_0 & F_{axis} > f_0 \end{cases} \quad (4)$$

where $f_0$ is the stalling force. The forces generated by the attached motor heads drive the motion of both the actin filaments and the motor aggregates. The motion of these elements is determined by calculating the forces acting on the nodes of the filaments and on the two ends of each motor cluster. The motion of all nodes (both of the filaments and motors) is treated as highly overdamped, i.e, a linear relationship is assumed between the instantaneous force and velocity: $v_i = F_i/\gamma$, with different drag coefficients $\gamma$ for the filaments and the motor aggregates. For the motor aggregate, a random Gaussian force ("white noise") is also introduced, whose magnitude is set so that the diffusion coefficient (along each Cartesian direction): $D = 2k_BT/\gamma$. Due to their large size, the diffusion of the filaments is ignored in the simulations. The position of each nodes is determined by $\Delta x_i = v_i \Delta t$, where $\Delta t$ is the time step of the MD simulations. Periodic boundary conditions were employed throughout the simulations. Generally speaking, the dynamics of the motor aggregates in our simulations is largely determined by the forces of the motor heads, whereas the diffusion mechanism makes only very little impact. Nevertheless, introducing white noise is essential to prevent disconnected motor aggregates from getting stuck in regions devoid of filaments. Since the objective of the work is to investigate the self-organization dynamics of filaments (to which distant disconnected motors do not contribute), and in order to reduce the computational toll associated with the simulations of "isolated" motor aggregates, we decided to accelerate the diffusion of such motors in a non-physical manner. This is achieved by increasing the amplitude of the random white noise for motors located a distance larger than some threshold from a filament, which causes all the motors to be drawn towards the filaments and essentially eliminates the existence of isolated motors.

The transition probabilities of motor heads between the attached and detached states are governed by the following rules: An unbound motor becomes connected to a filament with probability $p_{con}$ per time step if, and only if, the distance between them is smaller than $d_c$. The disconnecting probability per time step depends on the force $f$ exerted on the motor. It is given by $p_{\min-dis}$ at small forces $f < f_{dis}$; and at larger forces grows exponentially with the elastic energy stored in the Gaussian spring between the motor and the filament:

$$p_{dus} = \begin{cases} p_{\min-dis} & f \leq f_{dis} \\ p_{\min-dis} \exp\left((f^2 - f_{dis}^2)/2k_h k_B T\right) & f_{dis} < f < \sqrt{f_{dis}^2 - 2k_h k_B T \ln p_{\min-dis}} \\ 1 & f \geq \sqrt{f_{dis}^2 - 2k_h k_B T \ln p_{\min-dis}} \end{cases} \quad (5)$$

The model includes many parameters whose values have been determined from both physical and computational considerations. A detailed discussion of these issues, including a list of the model parameters and their chosen values can be found in the supporting information text 1 (see also ref. [30]). The results of our simulations are discussed and illustrated by figures in the following section. In the figures throughout the paper, each filament is represented as a black rod - see Fig. 1. The plus and minus end nodes are represented by a solid and an open circle, respectively. Motor aggregates, which in our model are represented by a rod with motor heads emerging from its both ends, are depicted as short red lines. The red line connects the ends of the rod, while the motor heads are not shown due to their very small size. Filaments and motor aggregates are drawn in the figures to the same scale.

**3. Results**

3.1 Local transport of two filaments

The forces driving the self-organization of active gels are generated by clusters of motor proteins that walk over the polar cytoskeleton filaments, bind them to each other, and lead to their relative movement. These clusters of non-processive motors tend to generate noisy dynamics, as is evident in supporting information (SI) movies 1 and 2 taken from the simulations and experiments, respectively. The irregular dynamics arise from the frequent changes in the forces exerted by the non-processive motors, which stochastically attach to and detach from the filaments causing the latter to move in a very discontinuous manner (see Fig. 2A). Over time-scales much larger than the characteristic binding time of motor heads, the motor clusters self-organize the system into mesoscopic patterns. Self-organization proceeds through several types of basic local configuration. The formation of these configurations stems from the tendency of motor heads to propagate towards the "plus" ends of the filaments which, in turn, causes two bound filaments to move relative to each other until their plus ends coincide. When the two filaments are aligned anti-parallel to each other, the motor heads connected to these two filaments are moving in opposite directions, as shown in Fig. 2B. When the two filaments are aligned in parallel directions (Fig. 2C), the motor heads propagate in the same direction. In the second case the filaments do not move until the plus end of one of them is reached. The motor heads then attempt to reach the plus end of the other, pulling it towards the plus end of the first one. A third possibility is that the two filaments are not aligned along a common axis. In this case, the relative orientation of the two filaments evolves from an "X configuration" (Fig. 2D) in which the two cross-linked filaments intersect, to a "T configuration" (Fig. 2E) in which the plus end of one of the filaments is brought into contact with the other filament. The final shape will be a "V configuration" (Fig. 2F), which is achieved once the motors reach the plus ends of both filaments. Over somewhat larger time scales, two filaments which are in the "V configuration" may rotate and align parallel to each other. This process resembles the closing of a "zipper". It is initiated by the motors that accumulate at the "V junction" and locally pull the filaments towards each other, effectively increasing the contact region between the filaments (Fig. 2G) and reducing the angle of the "V configuration". The process proceeds when new motors arrive at the contact region and cause its expansion until the two filaments become parallel. The formation of parallel arrangements of filaments was recently observed *in vitro* by Thoresen et al. [31]. It has also been shown that above a critical myosin density, bundles of parallel filaments are contractile. The formation of bundles and their contractility are discussed in the following subsection.

3.2 Bundles and asters

Over even longer periods of time, the system self-organizes into structures of multiply-connected filaments. The most abundant ones are "bundles" and "asters". Bundles consist of filaments that collectively align in both parallel and anti-parallel orientations. The properties of such active-filament bundles have been studied theoretically by several groups over the past decade. It has been argued that in such bundles, the dynamics depicted in Figs. 2B (Fig. 2C) during which anti-parallel (parallel) filaments are driven in the opposite (same) directions, would result in "polarity sorting" [12,32]. In the steady state of a fully sorted bundle, the filaments split into two groups, each of which consists of filaments with similar orientation (parallel filaments). The two groups, which are oriented anti-parallel to each other, are connected at their plus ends by the motors creating a structure resembling the one depicted in Fig. 2B for two anti-parallel filaments (but with two bundles rather than two filaments). This process is indeed seen in our simulation, although in many cases it fails to run to completion. In a partially completed polarity sorting processes, several bundle "subunits" with opposite orientations are formed, but at some point their progress in opposite direction is stalled and they are left with partial overlap. The origin of this phenomenon is probably the simultaneous binding of motors heads from the same end of the motor aggregate to several filaments whose plus ends are located at opposite sides of the attachment point. (The same scenario can, obviously, be encountered at both ends of the motor aggregate, which would aggravate the problem.) Because the individual motor heads attempt to move in opposite directions, the motor aggregate is stalled, and the motor heads detach from the filaments before making any substantial progress. Without such a progress, the filaments to which the motor heads are connected do not move either, and the whole polarity sorting process gets stuck. The result is a "dead-locked" bundle like the one shown, for example, in Fig. 3A. Notice in the figure the relatively high fraction of motor aggregates which are oriented perpendicular to the filaments. In this perpendicular configuration the motor heads do not exert forces that lead to relative sliding of the filaments.

Asters are "star like" structures made of filaments with their plus ends located at the center of the star. Such structures represent one of the common self-organized patterns in motor-filament systems. An aster is essentially a collection of "V configurations" (see Fig. 2F) of multiply connected filaments with a joint "core". The myosin II multi-headed aggregates serve as linkers that stabilize the aster structure. They accumulate at the "core" of the aster and bind to several "arms" simultaneously. Fig. 3B shows an example of an aster formed in our simulations. In this particular example, the aster has six arms radiating out from the aster core at equal angles from each other. Notice that each arm of the aster is a bundle consisting of several filaments. In our simulations, we usually observed asters with 3 to 7 arms, with clear preference to 4-arm asters with right angles between the arms. In fact, we noted that asters with a greater number of arms tended, over time, to transform into 4-arm ones, as demonstrated in Figs. 3C–3E. The merging of arms is driven by motors that propagate towards the core. Close to their destination, they manage to bind to two arms simultaneously and lead to their rotation in a manner resembling the "zippering" of a V-configuration shown in Fig. 2G. Our experimental results show that the angular distribution of the arms is rather homogeneous with many arms emanating from the aster core (see Fig. 3F, taken from our experiments). Asters with small number of arms have been observed in the presence of methyl cellulose which promotes the collapse of the arms into few thick bundles [10], but not in our experiments and other works. In our simulations, the collapse of arms is allowed due to the lack of excluded volume interactions between the filaments. Finally, we note that aster formation may be regarded as a two-dimensional polarity sorting process. As in the one-dimension case of bundles discussed in the previous paragraph, this process may not be fully accomplished resulting in, for instance, the aster shown in Fig. 3G. In this example, some of the arms exhibit "deadlocked" bundles with filaments whose plus ends fail to reach the core of the aster. The deadlocked arms can be easily identified by the smeared distribution of motors, which is marked contradiction with the fully sorted arms (Fig. 3B) along which only very few motors can be found.

3.3 Large scale organization

Thus far, our discussion has focused on the organization of small groups of filaments and motors. On such scales, the self-organization process is not influenced much by changes in parameters such as the length of the filaments and the concentrations of filaments and motors. These system parameters determine the structure on much larger scales. Fig. 4A shows the patterns formed after 16 seconds of simulations at various conditions. The "phase diagram" of patterns is plotted as a function of the length of the filaments (x-axis) and the number ratio of motor aggregates to filaments (y-axis). The figure shows that the system tends to evolve to one of the following two structures: (i) a collection of disconnected asters, or (ii) an interconnected network of filaments. The former usually appear for short filaments and at high concentration of motors, while the latter are generally formed for longer filaments and at lower motor concentrations. These trends can be rationalized as follows. Networks are formed by filaments that intersect with each other and become connected by motor aggregates at the intersection points. Their formation, therefore, greatly depends on the probability that the filaments cross each other, which obviously increases when the filaments are longer. Increasing the concentration of motors may have the opposite effect of disconnecting the networks into separated asters. This occurs due to the increase in the tensile stresses that the motors generate within the networks which, at sufficiently high concentrations, would rupture the network. The transition from network to asters with increasing the motor concentration has been also observed in our experiments (see Figs. 5A-C).

Another marked difference between systems of short and long filaments is the time needed for the system to reach the steady state. Fig. 4B shows snapshots of a system of short filaments at the initial state, after 4 sec, and after 16 sec. The steady state, consisting of a number of disconnected asters, is clearly observed already after 4 sec. By contrast, the system of long filaments Fig. 4C continues to evolve even after 16 sec. In this case, a network resembling the structure shown in Fig. 5A is formed after 4 sec of simulations. At this stage the system is organized into a fairly homogenous network of semi-flexible filaments slightly bent by the motor forces. After 16 sec, the network looks much less homogenous and exhibits longer and more flexible strands. These strands are formed when several filaments become connected by motors in a row into a fiber-like element (see sequence of snapshots Fig. 6A). Assembly of fibers has also been observed experimentally, as demonstrated in Fig. 6B. Sequential assembly of actin filaments and bundles to each other by myosin II motors is believed to be the mechanism by which stress fibers are generated in cells [33]. The fibers in our simulations and experiments are formed by essentially a similar mechanism. In stress fibers, the ends are anchored to focal adhesions, which provide mechanical resistance to actomyosin contractility. In our study, the fibers are free to slide, which is the reason that the system continues to evolve in time. Therefore, after very long durations (tens of seconds in simulations and a few minutes in experiments), the network either gets stuck in a mechanically meta-stable state or disintegrates into individual asters. The discrepancy between the time scales of the simulations and experiments can be attributed to several factors including: (i) possible inaccuracies in our estimation of values of the (many) model parameters (see SI - text 1), and (ii) the fact that the experimental system is denser and, moreover, includes excluded volume interactions slowing down the dynamics.

### 4. Discussion

In this paper, we presented a coarse-grained molecular simulation study of the self-organization of systems consisting of actin filaments and myosin II oligoclusters (intermediate-size aggregates of myosin II motors). The employment of a coarse-grained molecular description allows us to follow the dynamics from the molecular scales, where filaments are locally transported relative to each other, up to the macroscopically large scales, where the steady state structures are formed. We identify the different steps taking place during this hierarchical process, and clarify the role of the myosin II clusters as active linkers for the actin filaments. The observed dynamics and resulting patterns can be associated with the tendency of individual motor heads to progress towards the plus ends of actin filaments. Because the motor heads are grouped in oligoclusters they can bind simultaneously to several filaments, which result in the formation of structures with high connectivity junctions like bundles, asters, and networks. Our simulation results show a very nice agreement with experimental data from reconstituted actomyosin active gels. Most notably, both the experiments and simulations demonstrate how (for relatively long filaments and moderate motor densities) the system self organizes into quasi 2-dimensional networks which bear certain similarities to the structure of actin in the cell cortex. Another observation with direct relevance to cells is the ability of motors to generate fibers by inter-connecting overlapping filaments and bundles. A similar process may take place during the formation of contractile rings and stress fibers.

The ability of myosin II oligoclusters to induce formation of largely distinct structures is in contrast with passive actin binding proteins (such as α–actinin, fascin, etc.) which are usually involved in specific actin-based structures. This unique feature of the myosin is related to both: (i) its activity as a motor enabling rotation and transportation of actin filaments, and (ii) the association in oligoclusters consisting of a several motors that can simultaneously bind to few filaments. In a future work, we plan a more extensive experimental-computational study of similar systems, which would focus on the influence of the structure of the motor clusters. We believe that complex dynamics and a very rich phase diagram will be observed upon variations in different features of the motor clusters, such as the processivity of the motors, their number and spatial organization within the clusters. This will allow us to extend our studies in "two opposite directions" – to relatively small clusters with more processive motors (e.g., kinesin tetramers) on the one hand, and to very large myosin II aggregates (such as those found in muscle cells) on the other hand. We also plan to refine our model to include features like excluded volume effects, treadmilling of actin filaments and jamming of motors that propagate along the same filament.

The work was supported by the Israel Science Foundation (grant no. 1534/10).

# Figures

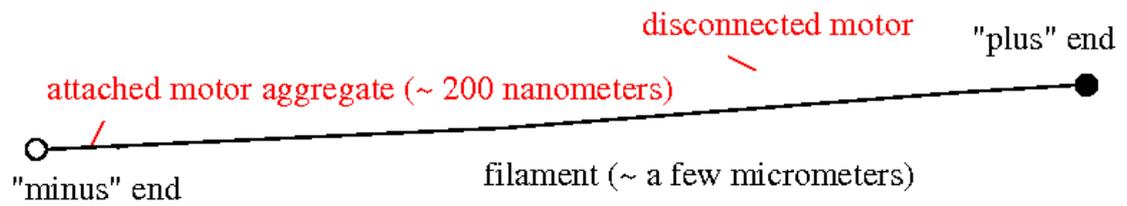

Figure 1: Representation of a filament (black line with a solid and an open circle at the ends) and motors (thin red line) in the figures throughout the paper.

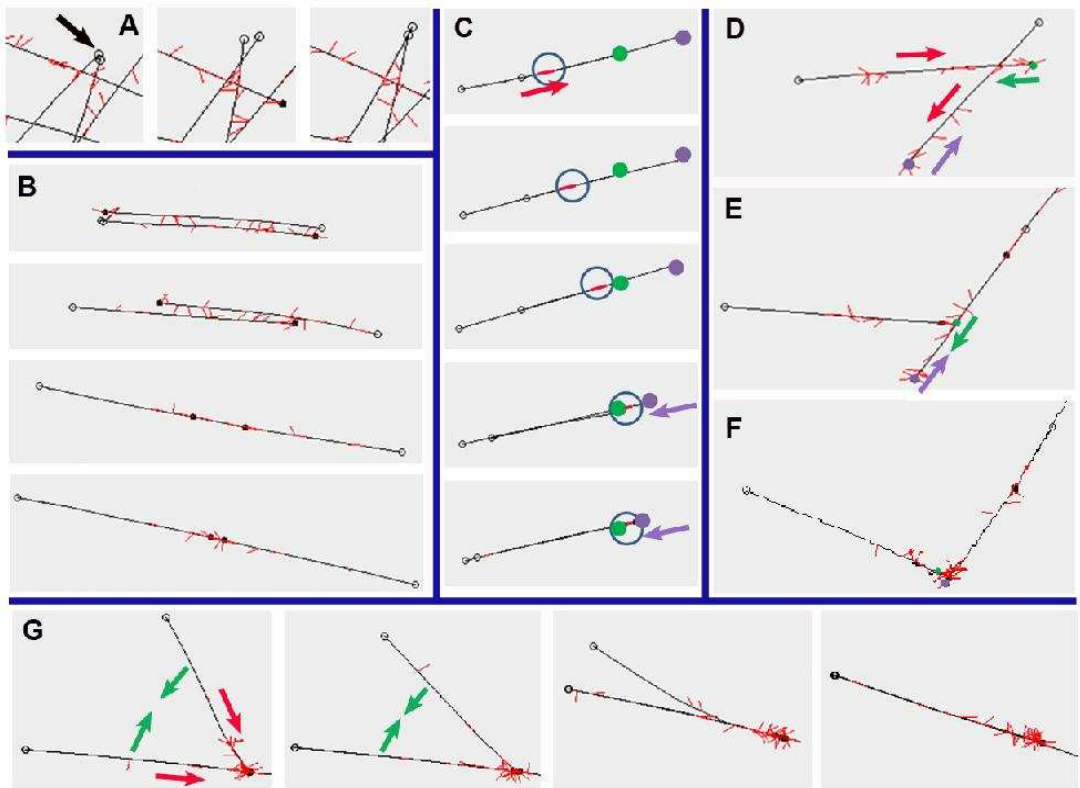

Figure 2: (A) Three sequential snapshots showing the dynamics of several filaments. In this sequence we see the discontinuous motion of two filaments whose minus ends (open circles indicated by an arrow) move away from and then back to each other. (B) The relative motion of two anti-parallel filaments. At the end of the process the two filaments overlap at their plus ends. (C) The relative motion of two parallel filaments. The plus ends of the two filaments are indicated by purple and green solid circles. In the first three snapshots, only the motors are moving (red arrow) along the filaments. Once the plus end of one of the filaments (green) is reached, the plus end of the other (purple) starts to move in its direction (purple arrow), until they coincide. (D-F) The evolution from the"X", through a "T", and into a "V" configuration. The red arrows indicate the motion of the motors, while the purple and green arrows show the motion of the filaments. (G) Transformation of a "V" configuration into a parallel bundle.
 

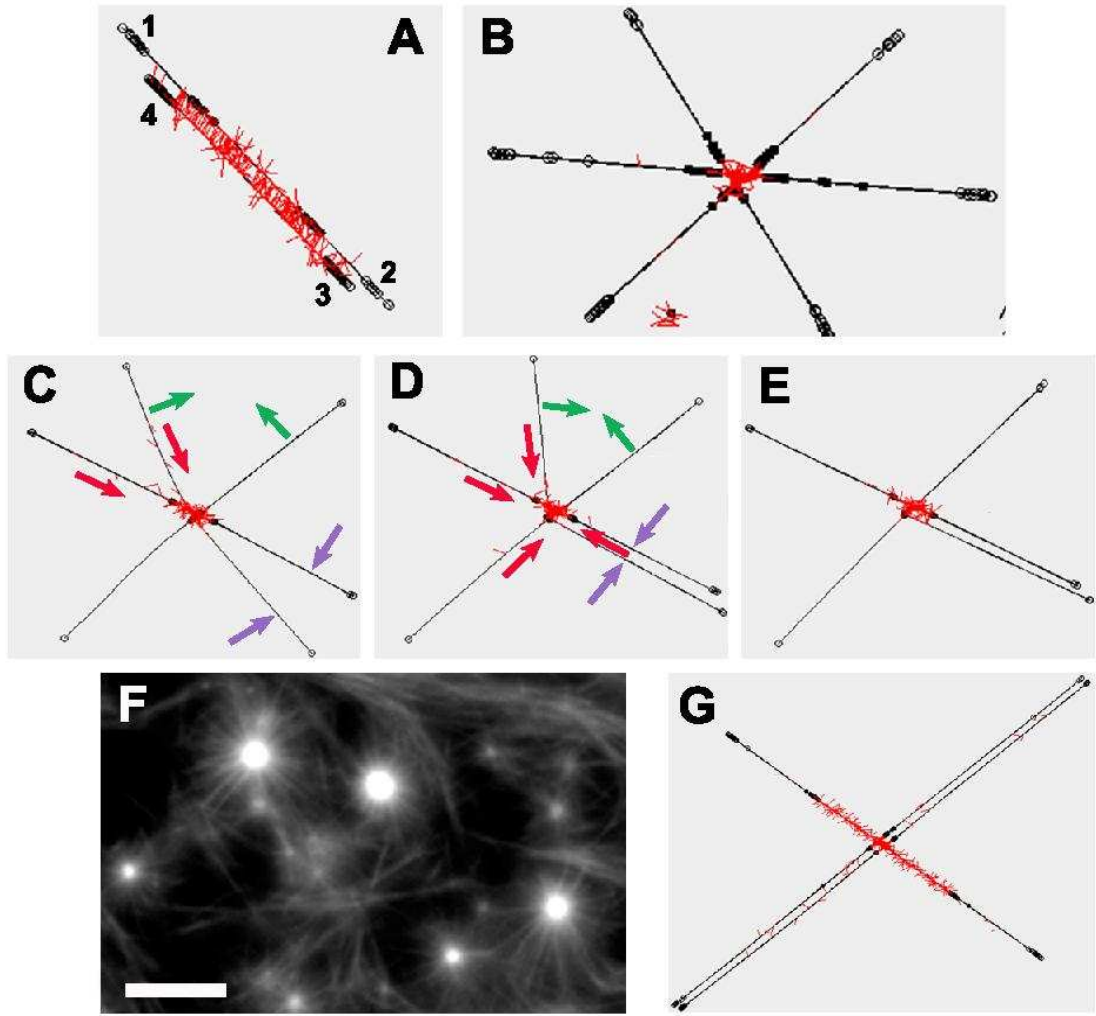

Figure 3: (A) Dead-locked bundle after partial "polarity sorting". In this specific example, the bundle consists of three groups of filaments whose minus ends are denoted by 1-4. (B) A typical 6-arm aster formed during the MD simulations. (C-E) The evolution of a 6-arm aster into an aster with 4 arms with right angles between them. (F) Asters formed in the motility assay. Unlike the simulations, these asters have many arms with a rather uniform angular distribution. Bar = 10μm. (G) A 4-arm aster with two fully sorted and two deadlocked arms. The later are characterized by a smeared distribution of motors, while in the former the motors localize mainly in the core.

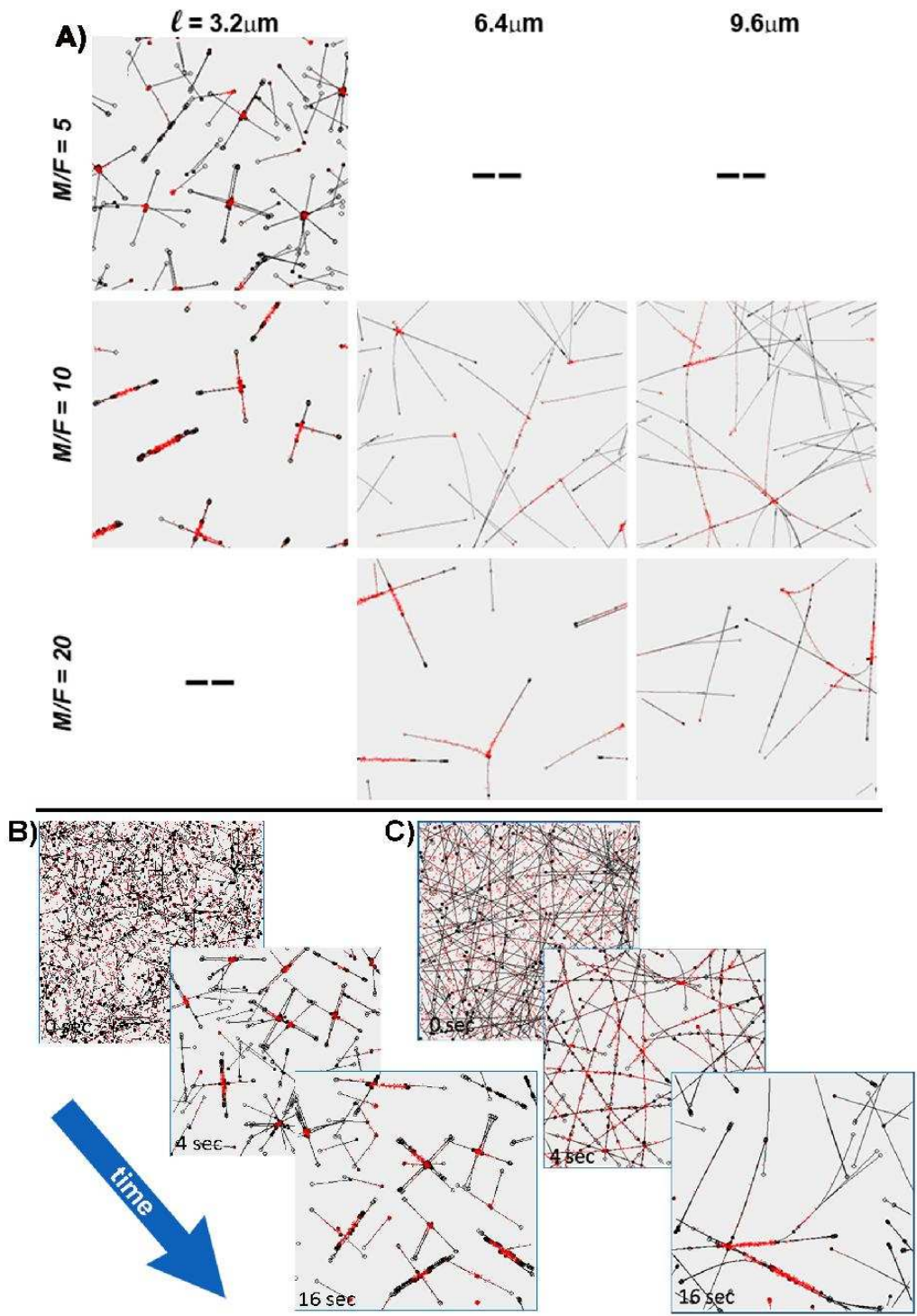

Figure 4: (A) Patterns formed after 16 seconds of simulations (starting from a random distribution of filaments and motors). Different patterns in the "phase diagram" correspond to systems with different length of the filaments $l$, and different motor to filament number ratio $M/F$. (B) Time evolution of a system of short filaments that form disconnected asters. (C) Time evolution of a system of long filaments that form a network. The total length of the filaments in 4B and 4C is the same. There are three times less filaments in 4C compared to 4B, and the length of each filament in 4C is three times longer than in 4B.

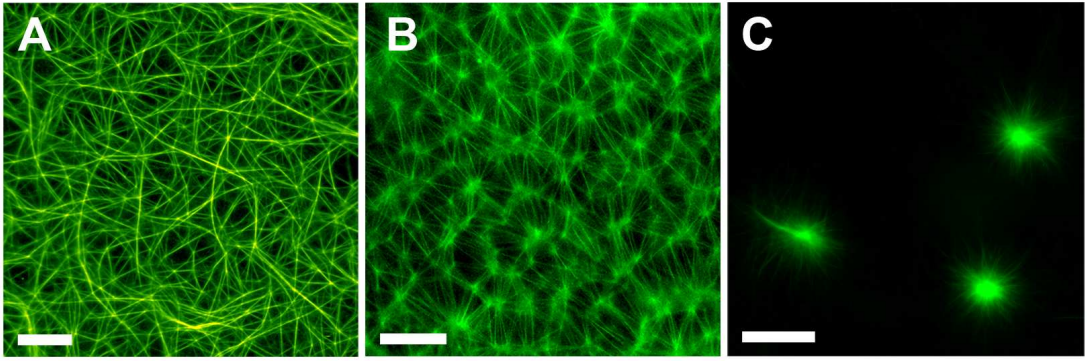

Figure 5: (A-C) Steady state structures of systems with different motor concentrations. At a low concentration of motors (A; concentration = 0.64 μM) we see a network of interconnected filaments. The formation of asters begins at intermediate concentrations (B; concentration = 1.0 μM), however at this concentration the asters are still connected to each other. Finally, at high motor concentrations (C; concentration = 2.8 μM), the network is ruptured and several distinct large asters are observed. Bar = 20μm.

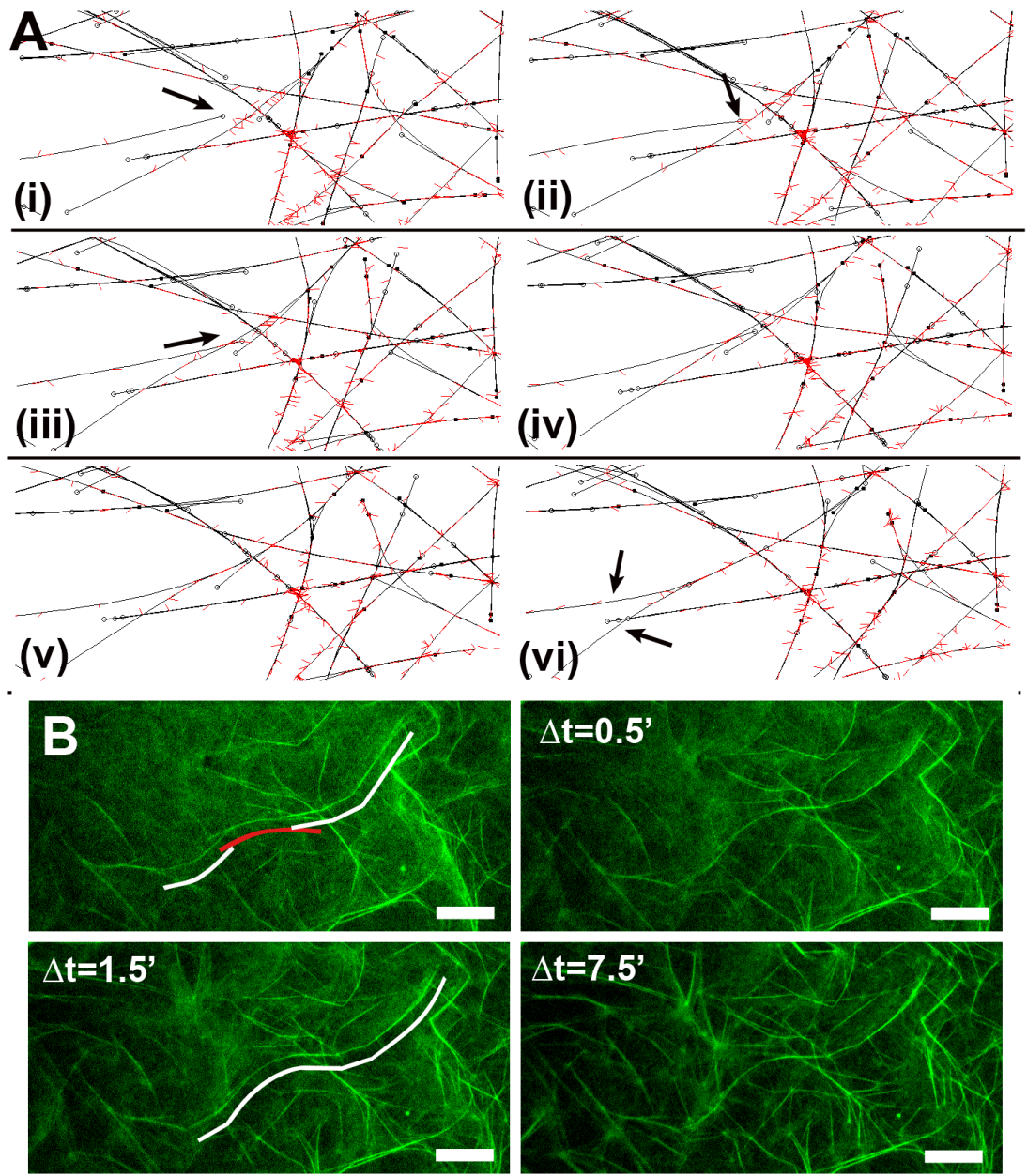

Figure 6: (A) A sequence of simulation snapshots showing how the end-to-end attachment of filaments (at positions indicated by arrows) leads to the formation of long fibers. The total time difference between (i) and (vi) is 0.25 sec. (B) Formation of a fiber from three filaments (white-red-white) in the experimental system. Concenrtaions: actin -15 μM, fascin - 2.14 μM, myosin - 1.5 μM. Bar = 20μm.